\documentclass[preprint,showpacs,preprintnumbers,amsmath,amssymb]{revtex4}
\usepackage{graphicx}
\usepackage{amsfonts}
\usepackage{amssymb}
\usepackage{amsmath}

\begin{document}

\title{%
How to generate and measure anomalous weakly non-ergodic Brownian motion in simple systems
}

\author{%
A. Fuli\'nski
}
\affiliation{%
M. Smoluchowski Institute of Physics, Jagiellonian University,
Reymonta 4, PL-30-059 
Krak\'ow, Poland
}

\begin{abstract}
It is shown that in systems with time-dependent and/or spatially nonuniform temperature $T(t,x)$, (i) most of the transport processes is weakly non-ergodic, and (ii) the diffusion (Brownian motion, BM) is anomalous. A few examples of simple arrangements, easy for experimental realization, are discussed in detail. Proposed measurements will enable also the observation of transitions from ergodic to weakly non-ergodic and from normal to anomalous diffusion. New effects are predicted: (i) 
zero-mean oscillations of $T(t)$ accelerate BM (pumping effect), (ii) the combination of temporal and spatial variations of temperature may lead to superballistic BM, (iii) linear gradients of $T(x)$ result in an exponential acceleration of BM. One can expect similar effects in inflationary systems with  time-dependent metrics.
\end{abstract}

\pacs{ 05.40.Jc, 02.50.Ey, 05.45.Tp}
 
\maketitle

In the recent literature \cite{{APP},{nonerg},{AF2011}} the process $X(t)$ is called {\it weakly non-ergodic} when:
\begin{equation}
\lim_{t_f\to\infty}\,
\langle \overline{ \delta^2X(t,t_f} \rangle \neq \langle X^2(t) \rangle\,,
\label{weak}
\end{equation}
(thorouhout this paper $X(t=0)=0$) with
\begin{equation}
\overline{ \delta^2X(t,t_f)} = \frac{1}{t_f - t} \int_0^{t_f-t}\! ds
\bigl[ X(s + t) - X(s) \bigr]^2
\label{avemov}
\end{equation}
relating two positions of the process (walker) separated by a time lag $t$
(for more details cf. \cite{AF2011}).

Weak ergodicity breaking (WEB) was discussed mainly in the context of anomalous diffusion (stochastic theory) \cite{nonerg}, though recently it was shown that WEB is typical for much wider class of cumulative-growth processes, both stochastic and deterministic \cite{AF2011}. Experimentally, WEB was found in composite disordered systems: various glasses, cytoplasm of living cells, and like \cite{disorder}. In this note we want to show how both anomalous diffusion (Brownian motion) and WEB can be generated and measured also in much simpler systems, easier to deal with and better suitable for various experimental manipulations.

Consider Brownian motion (BM) in a medium with variable temperature. The thermal (white Gaussian) noise $\xi(t)$ driving the Brownian particle is characterized by temperature-dependent intensity:
\begin{equation}
\langle \xi(t) \rangle = 0\,, \quad \langle \xi(t) \xi(s) \rangle = \sigma^2_{\xi}(T) \delta(t - s)\,.
\label{gwn}
\end {equation}

Consider first simple setups which can be easily realized experimentally: 
(i) thin (quasi-2d) flat layer (film) of a fluid which is cooling or heating (e. g. by emitting or absorbing radiation), (ii) long thin rod or fluid-filled cylinder immersed in a heat bath \cite{rem1}. When the characteristic time for equilibration of momenta is much shorter than the characteristic hydrodynamic (diffusional) time, it is safe to assume that the thermal local equilibrium is established with well-defined local temperature (in a coarse-grained sense). Note that when the cooling/heating is spatially uniform, there will appear no temperature gradients through the system, and we shall get the system in a quasi-adiabatically shifting thermal equilibrium.

In the considered system $\sigma^2_{\xi}$ will be changing in time through $T=T(t)$. For an ideal medium $\sigma^2_{\xi} \sim T$ (fluctuation-dissipation theorem), and we may write $\sigma^2_{\xi}(T) = T/T_0 \sigma^2_0(T_0)$, where $T_0 > 0$ is some reference temperature, e. g., $T(T=0)$; however, in real systems such dependence can be more complicated, depending on details of interparticle interactions. In a spatially non-uniform system the noise intensity will also depend on the (coarse-grained) position {\bf r}, $\sigma_{\xi} = \sigma_{\xi}({\bf r},t)$ (time-dependent diffusion coefficient was used recently for the description of random walks in porous materials \cite{ZZ}). 

The Brownian motion driven by the noise with $\sigma_{\xi} = \sigma_{\xi}(t)$ is weakly non-ergodic. Namely, consider kinetic (Langevin) equation \cite{LM}: 
\begin{equation}
\frac{d}{dt} {}_{s} X_{\beta}(t) = t^{\beta}\xi_0(t)\,,
\ \beta > -1/2\,,
\label{sbmL}
\end{equation}
where $\xi_0$ is the thermal noise at constant reference temperature $T_0$, with dispersion $\sigma_0^2(T_0)$. The formal solution of Eq.(\ref{sbmL}) is
\begin{equation}
{}_{s}X_{\beta}(t) =  \int_0^t u^{\beta} \xi_0(u) du \,.
\label{sbm}
\end{equation}
This {\it scaled Brownian motion} (SBM) is equivalent to the Brownian motion with time-dependent diffusion coefficient \cite{LM}: define the scaled noise $\eta(t) = t^{\beta} \xi_0(t)$. Then
\begin{equation}
\langle \eta(t) \rangle = 0\,, \ \langle \eta(t) \eta(s) \rangle = \sigma_{\eta}^2(t) \delta(t - s)\,, \ \sigma_{\eta}(t) = t^{\beta+1} \sigma_0\,,
\label{sgwn}
\end {equation}
and
\begin{equation}
\frac{d}{dt} {}_{s} X_{\beta}(t) =  \eta(t)\,.
\label{sbmL2}
\end{equation}

It was shown in \cite{AF2011} that SBM is weakly non-ergodic. This proof can be easily extended for any Brownian motion driven by thermal noise $\xi(t)$ with time-dependent intensity $\sigma_{\xi}(t) = \phi(t) \sigma_0$ (in fact this follows directly from the formulas (2.2)-(2.3) of \cite{AF2011}; for the sake of completeness, we present  here the specialized formulas). Namely, for 
\begin{equation}
X(t) = \int_0^t ds\, \phi(s) \xi_(s)\,,
\label{Xt}
\end{equation}
we get from Eqs. (\ref{weak}), (\ref{avemov}) and (\ref{Xt}):
\begin{equation}
\langle X^2(t) \rangle = 2\int_0^t\! ds\, D(s)\,, 
\label{xt1}
\end{equation}
\begin{equation}
2D(t) = \sigma^2_0 \phi^2(t) = \sigma^2_{\xi}(t,T(t))\,,
\label{DTt}
\end{equation}
\begin{eqnarray}
\langle \overline{\delta^2X(t,t_f)} \rangle &=& \frac{2}{t_f - t}
\int_0^{t_f-t} ds\, \int_s^{s+t} du\,D(u) \cr
\noalign{\vskip8pt} 
&&\hbox{\kern-58pt} =
\frac{1}{t_f - t} \int_0^{t_f-t} ds\,\bigl[ \langle X^2(s+t) \rangle
- \langle X^2(s) \rangle \bigr]\,,
\label{Xt2}
\end{eqnarray}
which is weakly non-ergodic except for $D=$ {\it const}, i. e., for the Wiener process (with constant temperature), and maybe for some special combinations of the spatial and temporal dependence of $D({\bf r},t)$ (i. e., $T({\bf r},t)$ -- cf. Eq.(\ref{scalingX}) below with $\gamma=2\beta$ for an example).

When the heating function $D(t)$ is proportional to $t^{\beta}$ (SBM case), this setup enables the realization of both WEB, and anomalous diffusion. Other forms of $\phi(t)$ will lead to non-diffusional (i. e., with dispersion not scaling algebraically with time) behavior of the (weakly nonergodic) Brownian motions.

Interesting is the case of periodic changes of temperature. Let $\phi(t) = 1 + A\sin(\omega t)$, $A^2 < 1$. Then 
\begin{equation}
\langle X^2(t) \rangle = 2D_At\, + 4D_0 \frac{A}{\omega} \bigl[1- \cos(\omega t) - \frac{A}{4}\,\sin(2\omega t)\bigr] \,, 
\label{sin}
\end{equation}
where $D_A = D_0(1 + A^2/2)$, $2D_0=\sigma_0^2(T_0)$, which tends asymptotically to normal diffusion with higher effective diffusion coefficient ({\it enhanced diffusion} or {\it accelerated Brownian motion}). The same effect will appear for anomalous Brownian motions, and for other forms of driving $\phi(t)$, either periodic or random. Note that when the asymptotic temporal average of the changes of temperature is zero, as in Eq.(\ref{sin}), we have a kind of {\it pumping} of energy into the Brownian particle, analogous to e. g. ionic nanopumping in asymmetric nanopores \cite{SF2002}.

More challenging for the analysis (though perhaps experimentally simpler) is the case when $T = T({\bf r},t)$. Analytical treatment is possible for the separable $T$, when $T({\bf r},t) = T_0\phi^2(t) \psi^2({\bf r})$, and only when the appropriate stochastic differential (Langevin) equation:
\begin{equation}
dX(t)/dt = \phi(t) \psi({\bf r}) \xi_0(t)
\label{Leqx}
\end{equation}
is treated according to the Stratonovich interpretation \cite{rem} (other situations need to be treated either by numerical simulations or by numerical solutions of the appropriate Smoluchowski-Fokker-Planck equations \cite{{VK},{ES}}). In that case, the Eq.(\ref{Leqx}) can be written in the form:
\begin{equation}
dY(t)/dt = \phi(t)\xi(t)\,, \ dY(t) = dX/\psi({\bf r})\,,
\label{Leqy}
\end{equation}
which means that the process $\{Y(t)\}$ is equivalent to the scaled Brownian motion, i. e., is weakly nonergodic \cite{AF2011}. To obtain properties of the original process $\{X(t)\}$, inversion of the transform (\ref{Leqy}) is necessary.
Note that for stationary systems, $\phi =$ {\it const.}, the process $\{Y(t)\}$ becomes the standard Wiener process $\{W(t)\}$, which is ergodic. This, however, does not imply that all original stationary processes $\{X(t)\}$ are also ergodic. Namely, the formal solution of Eq.(\ref{Leqx}) reads:
\begin{equation}
X(t) = \int_0^t ds\, \psi\bigl(X(s)\bigr) \phi(s) \xi(s)\,.
\label{sLeqx}
\end{equation}
Making use of the Stratonovich interpretation of stochastic integrals \cite{rem}, we get
\begin{equation}
\langle X^2(t) \rangle = 2D_0 \int_0^t ds \langle \psi^2\bigl(X(s)\bigr) \rangle \phi^2(s)\,,
\label{X2f}
\end{equation}
which is an integral equation for the dispersion of $X$.

Analogous calculations of the time average, Eq.(\ref{avemov}), give:
\begin{eqnarray}
\langle \overline{ \delta^2X(t,t_f)} \rangle &=& \frac{1}{t_f - t} \int_0^{t_f-t}\! ds \langle \bigl[ X(s + t) - X(s) \bigr]^2 \rangle 
\cr
\noalign{\vskip8pt} 
&&\hbox{\kern-58pt} =
\frac{2D_0}{t_f - t} \int_0^{t_f-t} ds\,\int_s^{s+t}\! du \langle \psi^2\bigl( X(u)\bigr) \rangle \phi^2(u)\,. 
\label{avemovX}
\end{eqnarray}
Thus, even for the stationary case $\phi = 1$, the process $\{X(t)\}$ is ergodic only when the second integral in Eq.(\ref{avemovX}) $\neq f(s)$, which is fulfilled for $\psi=$ {\it const.}, but besides maybe only for some very special choices of the temperature gradients. This proves that in most realistic cases the processes $\{X(t)\}$ in the media with spatial temperature gradients are weakly non-ergodic.

Further analysis depends on the forms of the functions $\phi(t)$ and -- especially -- $\psi({\bf r})$. Note that, when $D$ depends on {\bf r} and $t$ through its dependence on temperature: $D = D(T({\bf r},t))$, the position- and time-dependent temperature profile in a heated/cooled system has to be calculated from the Fourier-Kirchhoff equation (FKE):
\begin{equation}
\frac{\partial T}{\partial t} = \Lambda\,\frac{\partial^2 T}{\partial {\bf r}^2} + 
S({\bf r},t)\,,
\label{FKlaw}
\end{equation}
where $\Lambda$ is the heat conduction coefficient and $S$ -- heat source  \cite{rem2}. The solution of this equation for an arbitrary shape of the source $S$ can be found only by numerical computations, especially when we want to heat/cool the system in a very narrow region in the immediate neighbourhood of {\bf r} = 0. On the other hand, the FKE can be used for an inverse problem: one may calculate from Eq.(\ref{FKlaw}) the space- and time-dependent heat flow $S({\bf r},t)$ necessary for the attaining the prescribed temperature profile. In principle, any required $T({\bf r},t)$ can be used. However, while the calculation of the appropriate heat flow is a simple task, its experimental realization might not be so easy. 

As an example, consider a simple quasi-one-dimensional system: heat-conducting long narrow rod (or a fluid contained in a narrow cylinder) of length 2$L$ ($x \in [-L,+L]$, $L\to\infty$), with initial condition $T(X=0,t=0) = T_0 =$ {\it const}, and $X(t=0) = 0$. To retain the correspondence with anomalous diffusion, assume that $T =T_0\phi^2(t) \psi^2(X) = T_0(t + \tau)^{2\beta}(|X| + a)^{2\gamma}$, $\tau>0$, $a>0$. Then $a^{2\gamma}\,t^{2\beta}_0 = 1$, and $T$ is positive and bounded.

From Eq.(\ref{Leqy}):
\begin{eqnarray}
Y(t) = sgn( X(t))\begin{cases}
 \mu (|X| + a)^{1-\gamma}\,, \ \gamma\neq 1 \cr
\ln(|X(t)|/a + 1)\,, \ \gamma = 1\,. \cr
\end{cases}
\label{X2Y}
\end{eqnarray}
The inversion to $X(t)$ gives:
\begin{eqnarray}
X(t) = \begin{cases}
sgn(Y(t)/\mu) [\big| Y(t)/\mu \big|^{\mu} - a],, \ \gamma\neq 1 \cr
\noalign{\vskip5pt} 
sgn(X(t))a\bigl[ e^{Y(t)sgn(X(t))} - 1\bigr]\,, \ \gamma = 1 \cr
\end{cases}
\label{Y2X}
\end{eqnarray}
where $\mu = 1/(1-\gamma)$, and $sgn(f)$ denotes the sign of $f$ ($sgn(0) = 0$). Note that the initial condition $X(t=0)=0$ implies $Y(t=0)=0$.

The results for scaled Brownian motion mentioned above \cite{AF2011} give 
\begin{equation}
\langle Y^2(t) \rangle = 2D_{\beta}(t+\tau)^{\alpha_0}\,, \ \alpha_0 = 2\beta +1\,, 
\label{scalY1}
\end{equation}
and for $t_f\to\infty$, $t\ll t_f$
\begin{equation}
\langle \overline{\delta^2Y(t,t_f)} \rangle \to 2D_{\beta} (t+\tau)\, (t_f + \tau)^{\alpha_0-1} \,. 
\label{scalY2}
\end{equation}
where $D_{\beta} = D_0/\alpha_0$. Thus \cite{Gauss}, for $\gamma \neq 1$
\begin{eqnarray}
\langle X^2(t)\rangle &=& 2D_{\mu} (t+t_0)^{\alpha}\,, \cr
\noalign{\vskip8pt}
\alpha &=& \mu\alpha_0 = (2\beta +  1)/(1 - \gamma)\,, \cr
\noalign{\vskip8pt}
2D_{\mu} &=& (2D_{\beta}/\mu^2)^{\mu}\, \Gamma((2\mu+1)/2) \sqrt{2/\pi}\,, 
\label{scalingX}
\end{eqnarray}
and for $\gamma = 1$
\begin{eqnarray}
\langle X^2 \rangle &=& \exp\{4D_{\beta}(t+\tau)^{\alpha_0}\} - \cr 
\noalign{\vskip8pt}
&-& 2 a\exp\{D_{\beta}(t+\tau)^{\alpha_0} \} + a^2\,. 
\label{scaling1}
\end{eqnarray}
Note that $\gamma=1$ implies parabolic increase of temperature $T(X)$ with distance from the center.

On the other hand, direct calculations of $\overline{\delta^2X(t,t_f)} \rangle$ cannot be done analytically \cite{rem3}. However, to prove that the original process $\{X(t)\}$ is weakly non-ergodic, it is sufficient to prove that the ensemble and time averages of any function of this process are not equal. Therefore the result (\ref{scalY1})-(\ref{scalY2}) for the averages of $Y(t) = f(X(T))$ proves the weak ergodicity breaking of diffusional processes in systems with spatially nonuniform and/or time-dependent temperature $T(t,x)$. Note however that for $\gamma = -2\beta$ we get $\alpha = 1$, i. e., the process $\{X(t)\}$ becomes normal diffusion.

The results discussed in this paper show that the anomalous nonergodic transport is not restricted to composite systems, but can be easily generated in almost any system. Therefore, by (uniform) heating of the system with changes of temperature proportional to $t^{\beta}$ one can measure directly the weakly-nonergodic anomalous diffusion (both subdiffusion for $-1/2 <\beta < 0$, slowing-down the heating, and superdiffusion for $0 < \beta < 1/2$, accelerating the heating).
 
A corollary to this implies that the Mandelbrot-Van Ness (cf. \cite{AF2011}) fractional Brownian motion $B_H(t)$ which is ergodic, will become weakly non-ergodic in systems with varying temperature. Viz., the MBM of \cite{AF2011} can be written as $B_H$ driven by $\eta(t)$, Eq.(\ref{sgwn}):
\begin{eqnarray}
B_H(t) &=& \frac{1}{\Gamma(H+1/2)}\biggl[\int_{-\infty}^t (t-u)^{H-1/2} \eta(t;T(t))dt -
\cr
&-& \int_{-\infty}^0 (- u)^{H-1/2}\eta(t;T(t))dt \biggr] \cr
&=& \frac{1}{\Gamma(H+1/2)}\biggl[\int_{-\infty}^t (t-u)^{H-1/2} \phi(u) dW(u) -
\cr
&-& \int_{-\infty}^0 (-u)^{H-1/2} \phi(u) dW(u) \biggr]\,,
\label{fbm2}
\end{eqnarray}
where $dW(t) = \xi(t,T_0)dt$, and $W(t)$ is the Wiener process at temperature $T_0$.

Consider now a few specific situations, which lead to new, so far unknown effects, which will appear in systems with variable temperature.

(i) First such effect -- the enhancement of diffusion due to temperature oscillations was discussed above (Eq.(\ref{sin}))). 
The spatial analog of this situation, i. e., the stationary system with periodic (or aperiodic) variations of temperature, can be realized experimentally in the so-called {\it bistable ballast resistor}, i. e., thin conducting wire with nonlinear dependence of resistivity on temperature. When such systems are heated by electric current, and cooled by surface radiation, there appears the transition to the so-called dissipative structure with interlaced hot and cold fragments \cite{br}. Unfortunately, the analytic treatment of the Brownian motion in such situations (i. e., the inversion of the series $\{Y(t)\}$ into $\{X(t)\}$, and subsequent averaging over the noise) is practically impossible. 

(ii) Another effect is the superballistic Brownian motion resulting from the cooperation of the spatial and temporal changes of temperature. From the discussed above case of anomalous diffusion, when $\phi(t) = (t+\tau)^{\beta}$ and $\psi(X) = (|X|+a)^{\gamma}$ we find (cf. Eq.(\ref{scalingX})) that $\alpha >2$ when $\beta + \gamma > 1/2$.

(iii) The results described by Eq.(\ref{scalingX}) show that the Brownian motion accelerates exponentially in stationary systems with parabolic increase of temperature from the center towards the ends of a long (formally, infinitely long) cylinder. This means also that the logarithm of the dispersion of $X(t)$ behaves asymptotically as the normal diffusion. Linear $T(X) = T_0(|X| + a)$ results in the parabolic acceleration of the Brownian motion (ballistic motion).

As we have mentioned, the analytic treatment of the Brownian motion in systems with spatial temperature gradients is in most cases very difficult (or just impossible).
However, although the analytic calculations of dispersions (both ensemble and temporal) are too difficult,  it is possible to find the averages of some functions of $X(t)$, e. g. the inversions of the process $Y(t)$, or some functions of these inversions. 

All these assertions can be checked experimentally. This can be done by standard measurements of the time-dependence of the dispersion $\langle X^2(t) \rangle$ and -- when possible -- by the measurements of the single trajectories of Brownian particles. The latter would give direct experimental verification of non-ergodic character of SBM, and of other processes of this type. Note that the same can be done with other time characteristics of the speed of heating/cooling, including natural exponential cooling by radiation. Note also that it is important (and obvious) in such measurements that the time-scales of heating/cooling and of diffusion should be of the same order of magnitude.

It was shown here that anomalous diffusion (and related to it weak ergodicity breaking) can be detected in simple systems with nonstationary or/and nonuniform temperature. It should be noted that (i) similar effect can be found also in systems driven by external forces \cite{Des}, (ii) time-dependent diffusion coefficient was used recently for the description of random walks in porous materials \cite{ZZ}, (iii) one can expect similar effects in structures with  time-dependent metrics (inflationary systems).


\begin{thebibliography}{99}
\bibitem{Gauss}
Let us remind that $\{Y(t)\}$ is Gaussian-distributed; then $\langle \exp(\pm Y) \rangle = $ $\exp(\langle Y^2 \rangle/2)$, and $\langle |Y|^{\nu} \rangle =(2 \langle Y^2 \rangle^{\nu}/\pi)^{1/2}$ $\Gamma((\nu+1)/2)$.
\bibitem{rem3}
The comparison of the results for $\langle X^2(t) \rangle$ and $\langle Y^2(t) \rangle$ seems to suggest that also the average $\langle \overline{\delta^2X(t,t_f)} \rangle$ should behave similarly as $\langle \overline{\delta^2Y(t,t_f)} \rangle$, i. e., that it should scale as $(t+\tau)\, (t_f + \tau)^{\alpha-1}$.
\end{thebibliography}
\end{document}